\documentstyle[PASJadd]{PASJ95}
\draft
\input epsf
\epsfxsize=\hsize

\begin{document}
\setcounter{page}{1}

\title{Models for Accretion Disk Fluctuations through Self-Organized Criticality
Including Relativistic Effects}

\author{Ying {\sc Xiong}, Paul J.\  {\sc Wiita} and Gang  {\sc Bao}$^1$\\
{\it Department of Physics and Astronomy, Georgia State University,}\\
{\it University Plaza, Atlanta, Georgia 30303, USA} \\
$^1${\it current address: CIS Department, Columbia University,
525 W.\ 120th St., New
York, New York 10027, USA}\\
{\it E-mail(PJW): wiita@chara.gsu.edu}}  

\abst{The possibility that some of the observed X-ray and optical
 variability in active galactic nuclei and galactic black hole
candidates is produced in accretion disks through the development of a
self-organized critical state is reconsidered.  New simulations, including
more complete calculations of relativistic effects, do show that this
model can produce  light-curves and power-spectra for the 
variability which agree with the range observed in optical and X-ray studies 
of AGN and X-ray binaries.  However, the
universality of complete self-organized criticality is not
quite achieved.  This is 
mainly because the character 
of the variations depend quite substantially on the extent of the  unstable disk region.
If it extends close to the innermost stable orbit then a physical
scale is introduced and the scale-free character of self-organized
criticality is vitiated.  Significant dependence 
of the power spectrum density slope on 
the type of diffusion within the disk and a weaker dependence on
the amount of differential rotation
are noted. When general relativistic effects are incorporated
in the models, additional substantial differences are produced if the disk is viewed from
directions far from the accretion disk axis.
}

\kword{Accretion, accretion disks --- Black holes --- Galaxies: active ---
Hydrodynamics --- Relativity --- X-rays: binaries}

\maketitle
\thispagestyle{headings}

\section
{Introduction}

Structures within accretion disks can be observed indirectly
through the luminosity variations they engender. 
Several  hydrodynamical processes could produce such irregularities.
 Vortices
(e.g.\ Abramowicz et al.\ 1992; Bracco et al.\ 1999) and 
shocks induced by companion 
objects (e.g.\ Sawada, Matsuda, \& Hachisu 1986; 
Chakrabarti \& Wiita 1993) are among the
physical mechanisms that can yield significant
long-lived (multi-orbit) perturbations on disks.  

Any such quasi-coherent structures can produce 
``bright spots'' on the disk surfaces.  These regions of excess
emission, when coupled with the
disk's differential rotation and gravitational lensing of photons passing close
to the central black hole, can reproduce many aspects of the variations
seen in galactic black hole-candidates and active galactic nuclei (AGN).
The X-ray light curves and power spectra (Abramowicz et al.\ 1989, 1991; 
Zhang \& Bao 1991),
and optical/UV light curves and power spectra (Wiita et al.\ 1991; Mangalam
\& Wiita 1993) can both be understood within this bright spot model.
The anti-correlation between X-ray variability and luminosity in AGN, confirmed
by Lawrence \& Papadakis (1993), can also naturally be explained using
 this model (Bao \& Abramowicz 1996). It has also been 
demonstrated that the surprisingly strong coherence in the temporal
variability seen between different X-ray energies for galactic black hole
binaries (Vaughn \& Nowak 1997) is also capable of being understood within
the bright spot framework (Abramowicz et al.\ 1997).  This scenario also
predicts energy dependent polarization variations in the X-ray bands
(Bao, Wiita \& Hadrava 1996; Bao et al.\ 1997) which could be detected by
future X-ray satellites with polarimetric capabilities.  In addition,
changes in the X-ray spectra of Seyfert galaxies can be understood in
terms of varying amounts of reprocessing and reflection involving such
irregular accretion disks (Bao, Wiita \& Hadrava 1998; Bao \& Wiita 1999).

If accretion disks exhibit self-organized criticality (SOC),
as first suggested by Mineshige, Ouchi \& Nishimori (1994a), then 
another way to directly produce temporally extended quasi-coherent structures
is available.
  The SOC concept
arose from the realization that
a huge range of physical systems display  power-law or
scale-invariant correlations over decades in time or space.
``Flicker noise'' or ``1/f noise'' is seen in a
wide variety of systems, ranging from stock market prices, through
the height of the Nile river, to quasar light curves (e.g.\ Press 1978).
Bak, Tang \& Wiesenfeld (1987, 1988) showed extended 
systems with many metastable states can naturally evolve 
into a critical state with no fundamental length- or 
time-scales.  

In~\S\,2 we briefly summarize characteristics of the SOC state, and then 
 we mention some suggested astrophysical applications. In ~\S\,3 we 
discuss earlier efforts to show how this 
might work in accretion disks.  In~\S\,4 we display and discuss
additional
computations of modified SOC models for disks which display a goodly range
of light curves and variability power spectra. 
Some preliminary results (without, however, including relativistic effects)
were presented earlier (Wiita \& Xiong 1999; hereafter, WX). We find that the 
slopes of the power spectral density 
are quite sensitive to the extent of the region that is unstable, so that
these accretion disk models do not exhibit self-organized criticality
in the strictest sense. 
Our conclusions are given in \S 5.

\section{Characteristics of Self-Organized Criticality} 

\subsection{The sandpile model}

A sandpile was the first toy model for self-organized criticality,
and the original papers by Bak et al.\ (1987, 1988) performed numerical simulations
relevant to this situation, transformed into a cellular automata problem.  One
drops grains of sand randomly onto a pile until it builds up to a critical slope
or ``angle of repose''.  Once the pile is at SOC, the addition of a single extra
grain of sand anywhere causes the local slope to be too steep.
This leads to a small avalanche which readjusts the local slope to just below critical.
But now the moved sand will steepen the region of the pile 
just below it, perhaps making that region, in turn, too
steep.
Avalanches continue until the local slope everywhere adjusts to be at, or just below, 
the critical value.

The system becomes stationary when perturbations can just propagate the length of the
system, and a full-fledged SOC state of any physical system has the following characteristics: 
a distribution of minimally stable regions of all sizes;
small perturbations can yield ``avalanches'' of all sizes; 
the lack of any characteristic length scale produces a featureless
 power-law spectrum of avalanche sizes;
the critical state is insensitive to initial conditions.

Unsurprisingly, different physical systems have
different power-law indices for the relation between the number of events (avalanches) 
that occur with  different sizes or strengths.  These indices
depend on (Bak et al.\ 1988; O'Brien et al.\ 1991):  the number of spatial dimensions; 
the symmetry of the system;
and the attractor of the dynamics of the system.
For two-dimensional sandpile models, Bak et al.\ (1988) found the power spectral density
for the Fourier transform of the time-series to be $S(f) \propto f^{-1.57}$;  
for three-dimensional models they obtained 
$S(f) \propto f^{-1.08}$.

\subsection{Application to solar flares}

The first astronomical application of this concept
was the proposal by  Lu \& Hamilton (1991)  that the solar coronal magnetic 
field is in a SOC state.
They noted that observations of hard X-ray bursts in solar flares
have a distribution that is a power-law in peak photon flux with
logarithmic slope 1.8 over up to five decades:  $N(P) \propto P^{-1.8}$
(e.g. Dennis 1985). 

The Lu \& Hamilton model assumes that the
flares are avalanches of smaller magnetic reconnection events and that
all flares arise from the same physical process.
A lattice  picture in three dimensions assumes the local field
gradient is destroyed by reconnection when the local magnetic gradient  
is too large, as argued by Parker (1988).  
The magnetic gradient vector of a lattice point which exceeds this limit is eliminated
at that particular location, but the gradient is shared with neighboring lattice points.
In order for this model to work, changes in magnetic field must have a directionality 
(i.e., photospheric motions should on average increase $B$ in a 
particular direction, a not unreasonable assumption over substantial areas).

  The results of their simulations can be summarized
as follows:
energy release goes as $N(E) \propto E^{-1.4}$;
peak flux goes as $N(P) \propto P^{-1.8}$ (as observed).
Lu \& Hamilton (1991) argue that these power-law indices are rather insensitive to the 
critical field gradient.  While this picture is not exactly well  motivated
physically,  it does incorporate some viable assumptions
and does produce a way to understand an otherwise unexplained relation between the
 size and frequency of solar flares.  Additional physical conditions
as well as implications of 
this idea have been explored by Lu (1995).

\subsection{Application to Gamma-Ray Bursts}

Similar ``avalanches'' or ``chain reactions'' may play a
role in understanding details of gamma-ray bursts (Stern \& Svensson
1996).  While the time profiles of $\gamma$-ray bursts (GRBs) show
a very wide range of behaviors, there are several statistical properties that have
specific mathematical structures.  The average peak-aligned profile and the
autocorrelation function show stretched exponential behaviors.  This could imply that 
the wide range of GRB characteristics could arise
from different random realizations of a simple stochastic process which is scale 
invariant in time.

The Stern \& Svensson (1996) model is based upon a pulse avalanche, where a trigger event
induces many additional flares by setting off a ``chain reaction'' in a near-critical
regime.  Such a basic picture was shown to fit both the diversity of GRB time-profiles
and important average statistical properties of GRBs, such as their third-moment, 
their autocorrelation function and their duration distribution function.  
Similarly to Lu \& Hamilton
(1991), Stern \& Svensson (1996) suggest that magnetic reconnection in a turbulent
medium is a plausible underlying mechanism, one that can fit within 
exploding fireball or relativistic jet models for GRBs.

\section{SOC and  Accretion Disks}

The first proposal that accretion disks around black holes could be in an SOC state
was made by Mineshige, Ouchi \& Nishimori (1994a).  This was an attempt to produce 
the steeper than $\sim 1/f$ fluctuations seen
in X-ray power-spectrum densities (PSD) 
for BH candidates such as Cyg X-1 (e.g.\ Miyamoto et al.\ 1993).
This could provide physical connections between bright spots (or determine
the size thereof) and thereby engender coherent structures for interesting
lengths of time.  Our work in this paper is essentially an extension and 
modification of this pioneering effort, so we shall describe their approach in moderate
detail.  An extensive review and discussion of other aspects of
this approach has been given recently by Mineshige \& Negoro
(1999).

The outer disk is assumed to be smooth enough so as to not produce any variability; 
however, the inner disk is subject to some, not yet physically constrained,
instability.  (Mineshige et al.\ suggest this instability could be
related to flares in the corona, 
presumed to lie above and below the disk, in
analogy with Lu \& Hamilton's solar corona picture.) 
The unstable inner disk is divided into zones ($i,j$) characterized
by radial ($r_i$) 
and azimuthal ($\phi_j$) coordinates and taken to be geometrically thin.
To model ordinary viscous accretion, the outer disk feeds a mass particle, $m$, to a zone 
in the outermost ring of the inner disk, $r_1$, with random $\phi$.
In lieu of more developed physics, Mineshige et al.\ (1994a) assume 
$M_{{\rm crit},i} \propto r$. If the mass
contained in the $(i,j)^{th}$ zone, $M_{i,j} >
M_{{\rm crit},i}$, they then assume enough mass is dumped from zone ($i,j$) to bring $M_{i,j} \le
M_{{\rm crit},i}$.  This excreted mass is taken to spread to the three
nearest  zones one ring interior to the destabilized zone 
(for simulations corresponding to rigid rotation) or to 
three trailing interior zones (as a simple way of mimicking differential rotation).

After the mass has been dumped from an overloaded zone into others closer to the
central mass, one must test those adjacent interior zones for $M_{i+1,(j-1, j,
j+1)} > M_{{\rm crit},i+1}$ (these choices for the second subscript pertain 
to rigid rotation).  
If this condition is fulfilled, one continues the avalanche by dumping mass into next 
ring inwards.
They reasonably estimate the luminosity for each dumped $m$ by:
\begin{equation}
L_x \simeq  (GM_{\rm BH}m)\Bigl({1\over r_f} - {1\over r_i}\Bigr),
\end{equation}
where $M_{\rm BH}$ is the mass of the central black hole, and $r_i$ and $r_f$
are the initial and final radii experienced by that particular mass unit
during that particular flow inwards.
Then one sums up for the energy released by all avalanching masses.
Finally, one reinjects mass at the outer edge of the inner disk and repeats
the above procedures.
For their canonical  disk model, Mineshige et al.\ (1994a) discover approximate scalings
where the number of 
events vary with their energy as
$N(E) \propto E^{-1.35}$, the lifetime distribution varies as $N(\tau) \propto
\tau^{-1.7}$, and the PSD follows $S(f) \propto f^{-1.8}$.  They showed these power-law
indices, or slopes,
to be nearly independent of the amount of differential rotation artificially implemented
in their simulations.

This type of scenario was also proposed by Yonehara, Mineshige \&
Welsh (1997) as a way to explain the PSD of optical and ultraviolet
fluctuations in cataclysmic variables.

\subsection{Diffusion Plus Avalanches in Disks} 

The above scenario was modified by Mineshige, Takeuchi \& Nishimori (1994b), 
who included
gradual mass inflow in every timestep, even if all zones were stable, 
with $M_{i,j} < M_{{\rm crit},i}$.  By doing  so Mineshige et al.\ (1994b) were able to model 
continuous accretion through viscosity in the disk, even
in the absence of substantial mass flow (and energy release) produced through instabilities.
In their new standard model they transfered $3m$ if a zone is unstable (as before) and 
also $0.01m$ from one randomly chosen cell in each ring during each fundamental timestep.

The key result was that the PSD index $\beta$ for  $S(f) \propto f^{-\beta}$ was reduced
from $\sim$1.8 to $\sim$1.6.
They note that this is in better agreement with  AGN 
(e.g.\ Lawrence \& Papadakis 1993) and galactic BH candidate data 
(e.g.\ Miyamoto et al.\ 1993).
Raising the diffusion rate substantially lowers $\beta$ somewhat further.
Other results from these models are that diffusion produces  more small-scale
avalanches in the inner regions, and decreases the chance for bigger flares.
Small shots (energy releases) usually appear randomly in time since the mass is 
deposited into the outer ring at random points.
On the other hand, big shots could have a non-random temporal 
distribution because a big event drains many cells below their critical values and 
therefore a longer time is needed to build back up to the critical state.  Careful studies
of the temporal distribution of flares of various sizes could therefore, in principle,
provide a test of this SOC-type model.  An exponential distribution
of peak intensities arises from this model and provides a better
fit to some of the data (Takeuchi, Mineshige \& Negoro 1995).

\subsection{Extrinsic, relativistic modifications}

Preliminary and quite approximate considerations of general relativistic effects, 
primarily Doppler boosts and gravitational light bending, were also shown
to lower $\beta$ (Abramowicz \& Bao 1994).  This work combined the
basic idea of the bright-spot model (\S\ 1) with the SOC concept, but involved
significant analytical approximations. The extent to which 
these extrinsic effects modify the PSD index 
 is a strong function of the viewing
angle, $i$, to the normal to the disk plane in this approximation. Larger values 
of $i$ were found to produce shallower slopes,
with $\beta$ reduced to $\sim$1.4  for $i = 80^{\circ}$ from $\sim$1.8 for $i = 0^{\circ}$
(Abramowicz \& Bao 1994).

\subsection{Application to Advection Dominated Accretion Flows}

This model was further modified by Takeuchi \& Mineshige (1997)
to incorporate key features of the
advection dominated accretion flows, which were
 first suggested by Ichimaru
(1977) and since invoked by many others (e.g.\ Chen et al.\ 1995).  
They pointed out that such flows could provide an
even better substrate in terms of satisfying several details
of observations of galactic BH candidates.  The basic nature of
the X-ray shot was argued to be well fit by a disturbance
propagating inward through an advection dominated disk
(Manmoto et al.\ 1996) and  Takeuchi \& Mineshige (1997)
built upon this concept.  They considered
various values of the critical radius at which the surface
density became too high, and showed, within the framework
of a time-dependent but one-dimensional calculation, that flatter
PSDs were obtained for larger values of this critical radius.

\section{Extended Models  for SOC  in Accretion Disks}

Our new efforts have involved making changes to various parameters for models similar
to those suggested by Mineshige et al.\ (1994a,b) to see if the light curves and
power spectra for the fluctuations were actually as independent of initial and
boundary conditions as they suggested, particularly when general relativistic
effects were included.  We must stress that all of our work to date has been 
restricted to non-rotating black holes. We examined models with different values of the:
\begin{itemize}
\item inner and outer radii of the unstable inner portion of the disk;
\item viewing angle;
\item amount of mass dumped when a cell exceeded the stability limit;
\item amount of differential rotation in the location of the zones where the mass is
dumped;
\item diffusion to sideways and outward as well as inward zones;
\item initial deviations from critical density;
\item accretion rate. 
\end{itemize}

\subsection {Results excluding relativistic effects}

Some preliminary results, not including proper relativistic effects
and therefore only able to consider face-on views of the disks, 
were given in WX.
As a first step towards inclusion of relativistic effects, we have also
modified the gravitational potential from the Newtonian, with $\Phi \propto r^{-1}$, 
to the pseudo-Newtonian
type, $\Phi \propto (r-2)^{-1}$; the latter ``Paczy\'nski -- Wiita potential''
 quite accurately reproduces many of the 
leading effects of the Schwarzschild geometry  (Paczy\'nski \& Wiita 1980; 
Artimova, Bj\"ornsson \& Novikov 1996).  With the above expressions we begin the use
of notation where distances are given in units of $r_g = GM_{BH}/c^2$.
Then the energy released is,
\begin{equation}
L_x \simeq  (GM_{\rm BH}m)\Bigl({1\over (r_f - 2)} - {1\over (r_i - 2)}\Bigr).
\end{equation}

We begin by summarizing our first simulations, which did not include
relativistic effects (these were discussed in more detail in WX).  
These early simulations  used  the standard
model of Mineshige et al.\ (1994a), with 64 rings of unit radius extending inward 
from $r_1 = 100$
and with the individual zones starting just slightly below critical densities; given
those parameters
we do indeed reproduce a quasi-power law PSD with slope 
$\beta \approx 1.8$.
We then considered simulations where the radial extent of each ring was modified
to produce equal intervals in the logarithm between specified inner and outer
radii, and found that for the situation considered by Mineshige et al.\ (1994a),
differences between uniform and logarithmic gridding were negligible, so we
have adopted the logarithmic scaling in our work.  This allows for smaller unit
regions of emission closer to the central BH, which is 
 more realistic as larger gradients of the gravitational field exist
in regions close to the black hole; it also better
tracks the increase in density and magnetic field energy expected in inner
disk regions.  In all of our work, the amount of mass dumped from each zone
that is triggered
by being pushed over its critical value is assumed to be constant, and
is denoted by $m_d$, despite
the varying sizes of the zones.

These PSD slopes depend only 
very weakly upon most of the parameters needed to characterize the simple model.  
Unsurprisingly,  the
early stages of the light curves are very sensitive to the initial deviations from the 
critical density.   Only very small fluctuations are seen at 
early times  if the dumped mass satisfies the inequality, $m_{\rm d}/3 <
M_{\rm crit} - M_{\rm init} \equiv \Delta M$; however, the disks eventually approach the 
critical values in most zones, and the later portions of the light-curves appear
 similar to each other. The stable portions of the
 PSDs (those pertaining to higher frequencies) have very similar slopes even if there is 
a substantial delay before the  critical state is reached throughout the disk.
Thus we do not display here results involving modifications to the initial deviation
from critical density.

Low values of the accretion rate tend to yield larger numbers of flares of specific
energies, since, at early times, only single mass dumps or small multiples of these
fixed mass dumps (and hence energy production) are induced.  In such cases
 the light curves can retain a banded (or ``quantized'') appearance for significant
times (see Fig.\ 3 of WX); nonetheless, values of $\beta$ are quite independent of $\dot M$,
once ``avalanches'' of wider ranges occur,  and an at least quasi-SOC situation is established.

In our non-relativistic simulations we also found the following effects,
which we will discuss more completely in \S 4.2, since they were also considered using
our relativistic code.  In modeling differential rotation, both the constant extra zone
shifts considered by Mineshige et al. (1994a) and variable zone shifts, which better
approximate some effects of the actual Keplerian rotation curves, were considered.  
But in all cases, we found only
small changes in light curves or slopes of the PSDs ($\Delta \beta < 0.04$), in agreement
with  Mineshige et al. (1994a).  Further, 
in agreement with Mineshige et al.\ (1994b) we found more  dependence of the results
on the nature of the diffusion. They showed that if there is extra diffusion inward, then 
the slope of the PSD becomes
shallower.  This additional diffusion is a reasonable approximation to standard viscosity in disks, which
transports matter inwards and angular momentum outwards.  We also performed simulations
which allow for a more general diffusion of energy, which might be more realistic if
actual magnetic reconnection events do indeed provide the physical trigger. To allow for
propagation of a disturbance partially sideways and back outwards as well as inwards 
(which was still presumed to
dominate) we also considered cases where the mass outflow from an unstable zone
would go 1/5 to each of three interior zones (thus 60\% inwards), 1/8 to each of
the two neighboring zones in the same ring (25\% sideways) and 1/20 to each of the three 
exterior zones (15\% outwards).
We find that if the triggering zone can send some matter back outwards and sideways,  
then the PSD slope
can become steeper by up to $\Delta \beta \approx 0.15$--$0.25$ in the non-relativistic
approximation (cf. Fig.\ 4 of WX). 

For the non-relativistic simulations we found a
substantial sensitivity of the power spectrum density index $\beta$ to the location
of the disk's inner radius ($r_N$) with respect to its outer radius ($r_1$).  
Even without including diffusion, $1.1 < \beta < 1.8$, was found, with the shallowest slopes
corresponding to $r_N \approx 6$, i.e., at the innermost stable orbit for a non-rotating 
black hole, when the outer radius is fixed at $r_1 = 200$.
Some modest changes are induced by changing the energy released in Eq.\ (1) from a pure
 Newtonian,
and nominally scale free, $r^{-1}$ dependence, to the pseudo-Newtonian, 
$(r-2)^{-1}$, dependence of Eq.\ (2).
In cases where the inner radius was set at $r_N = 6$, so this effect 
would be maximized, we found the slope to flatten by $\Delta \beta \approx 0.1$.

\subsection{Results of Relativistic Simulations}

In that most accretion disks will be viewed at some angle other than face-on
($i = 0^{\circ}$), it is important to consider what effect alternate viewing angles could produce
on the observed light curves and PSDs.  As mentioned in \S 3.2, Abramowicz \& Bao (1994) incorporated
an approximate treatment of such effects and noted that significant flattening of the
PSD slope could be expected.  In our current work, we have used a more thorough treatment
of relativistic effects on the propagation of energy emitted in the vicinity of a Schwarzschild
black hole.  

Our calculations retain the basic scenario of Mineshige et al.\ (1994a)
as modified by WX; the basic procedure is discussed above and
described in more detail in Mineshige et al. (1994a, b)
and Mineshige \& Negoro (1999).  But the results presented here
incorporate both the pseudo-Newtonian 
potential to describe energy release
(approximate, but reasonable, given the phenomenological nature of the instability model)
as well as full general relativistic (GR) effects on the propagation of the emerging
radiation.  Note that the light curves presented below exaggerate
the magnitude of the fluctuations in that no effectively
quiescent underlying disk emission has been included.  However,
the preemptive subtraction of this mean level does not affect
the PSDs.

These GR effects on propagation consist of light-travel 
time differences, gravitational redshifts, Doppler boosting, 
and near-field gravitational lensing, and are computed using the
efficient techniques described in Bao, Hadrava \& Ostgaard (1994)
and Bao et al.\ (1997); we refer the reader to these works and
do not repeat the details of these techniques here.  For a given 
location of an element
$(r, \phi)$, and the viewing angle of the observer ($i$), we can easily integrate
the general relativistic effects on photons emitted by the
element into the observed light intensity. 
  We note that these
techniques have been generalized to predict the nature of
 polarization variability from
unsteady accretion disks (Bao et al. 1996,
1997) and have been successfully applied to emission from radio jets 
(Bao \& Wiita 1997) 
and to
the X-ray variability from Seyfert galaxies (Bao, Wiita \& Hadrava 1998; Bao
\& Wiita 1999).  

The radiation is assumed to emerge uniformly from each
grid zone.  These zones have an essentially Keplerian 
azimuthal velocity taken to vary as (Paczy\'nski \& Wiita 1980)
\begin{equation}
v_{\phi}(r)  \propto \Bigl( \frac{GM_{BH}}{r} \Bigr)^{1/2} \Bigl( \frac{r}{r-2}
\Bigl),
\end{equation}
and are normalized to have $0.5c$ at $r = 6$, which is exact
for standard thin accretion disks (e.g.\ Shakura \& Sunyaev 1973).
For such thin disks the radial component of the flow, $v_r \ll
v_{\phi}$, except for very close to $r = 6$,
and thus $v_r$ can be neglected in our calculations.
The lifetime of the emission from each triggered zone is taken to
be 8 times the local orbital period; while somewhat arbitrary, this
assumption allows a full range of variations to build up.
  The simulations are run for extended periods, with the results
reported here based on taking either $4 \times 10^5$ or $1 \times 10^6$
timesteps.  Each timestep corresponds to 1/64 of the orbital period
at the inner edge of the unstable disk region, so that 
thousands of orbits (even at the outer edge
of the unstable disk region) are considered.  Thus, if an SOC state were to
be established, these simulations are lengthy enough to discern it
(e.g.\ Mineshige \& Negoro 1999).
Aside from our not considering Kerr geometries, one other geometrical
limitation of these models must be stressed: the accretion disk is assumed to be
completely thin and flat.  The PSDs are determined from
the ``observed'' light curves, i.e., after GR effects on propagation
are included, using {\sl Mathematica}.  The slopes of these
PSDs, $\beta$ are computed from the dominant region of the PSD; on the
scale-free PSDs presented below, this region is defined as extending over
1.6 decades in frequency, from 50 to 2048.

Table 1 summarizes the parameters of the 43 different simulations we have
performed and analyzed in this current work.  All of these include the relativistic
effects discussed above. The column headings identify the Case, give the
inner ($r_{in} \equiv r_N$) and outer ($r_{out} \equiv r_1$) radii, the viewing angle, the type of rotational
prescription considered, the type of diffusion prescription considered,
and the amount of mass dumped when a zone goes unstable; the summary output
parameter is the slope of the PSD in the region at high enough frequency that
an approximate power-law is established.  In several cases, 
multiple realizations of the same parameters have been averaged;
however, since the computed
PSD slopes, $\beta$, never varied by more than 0.02 between such
 different realizations,
in general only single realizations of these simulations are 
used to produce the values of $\beta$ quoted in Table 1.

Cases 1--9 illustrate the situation for disks where only a very limited inner region, 
from $r_N = 6$ to $r_1 = 19$, is unstable.  Such a situation is expected in
many standard thin accretion disk models where radiation pressure dominates the
innermost portions (e.g.\ Lightman \& Eardley 1974).
We also note that transonic accretion flows yield shocks at comparable
radii, and the flow within these regions is more unstable
(e.g.\ Chakrabarti 1996).
The differences between the individual Cases in this first group involve 
changing the viewing angle from face-on ($i = 0^{\circ}$), where focusing and light-travel delays
due to GR are negligible, though red-shifts and resulting 
reductions in observed power from the
innermost regions still play a small role,
to nearly along the disk ($i = 85^{\circ}$), where gravitational focusing can
strongly amplify energy releases emitted on the opposite side of the disk
from the observer (e.g. Bao et al.\ 1998).  A light-curve and PSD for Case 2
are displayed in Fig.\ 1.  The entries in Table 1 for Cases 1--6 show
 that the PSD flattens as the inclination
angle rises from $0^{\circ}$ up through $60^{\circ}$, with a total change of
$\Delta \beta \simeq 0.35$, in rough accord with the more approximate
calculations of Abramowicz \& Bao (1994).  As the inclination rises, the 
effects of Doppler shifts 
act to enhance effects of energy avalanches extending close to the BH
since the velocities there are higher.  This raises the power at higher
frequencies with respect to those at lower frequencies,
thereby flattening the PSD. 
As the viewing angle increases still further, near-field gravitational 
lensing becomes more important and the
PSD steepens again for $i > 70^{\circ}$.  This effect was not
found in the more approximate treatment of Abramowicz \& Bao (1994).

In the subsequent sets of simulations we explore situations where larger regions of
the disk are taken to be unstable.  For Cases 10--18 the inner boundary remains at the last
stable circular orbit, but the outer radius is extended to $32(GM_{BH}/c^2)$.  Cases
19--27 again have the same inner boundary, but the outer boundary is taken to 
go all the way out to 100.  The key result of these simulations is that,
by including larger regions of instability, the slope of the PSD flattens considerably.
Figure 2 gives a light curve and PSD for Case 11, and Fig.\ 3 is for Case 20;
these have the same inclination (and all other properties) as did Case 2, illustrated
in Fig.\ 1.  The fundamental trend of producing substantially flatter PSDs
from larger unstable
regions is very clear, with $\Delta \beta \sim 0.10 - 0.25$ between $R_1 = 19$
and $R_1 = 32$ (the extent of the change in slope varies with viewing angle), and 
$\Delta \beta \sim 0.7 - 0.8$ between $R_1 = 19$ and $R_1 = 100$.  Trends
with viewing angles, for fixed boundary radii, are similar to those seen in Cases 1--9: 
as $i$ rises from $0^{\circ}$ to higher angles ($70^{\circ}$
for $R_1 = 32$; $80^{\circ}$ for $R_1 = 100$) the value of $\beta$ drops by $\sim 0.25$
from the value for face-on observations, demonstrating how the
Doppler boosts become more important.  But, at the very highest angles, the near-field lensing
plays a dominant role, partially reversing the flattening.
Unsurprisingly,  the greater the outward extent of the unstable region, the lower the relative
importance of the latter effect, and the ``turn-around'' in the 
slope of the PSD begins
at higher angles and is less pronounced.  Figure 4 shows 
some of this
particular variation, displaying Case 13, which has a higher inclination, but is
otherwise identical to Case 11 (Fig.\ 2).

We also vary the boundaries of the unstable region in our next set of simulations, Cases 28--32, 
but we do so by moving the innermost radius of instability to 20 instead of 6.  
These would be expected
to have steeper PSDs for outer boundaries of the same value (100, e.g.\ Cases 19--27),
because the fastest moving zones at the smallest radii, which contribute more power at the highest
frequencies, are no longer included.  Case 28, with $i = 0^{\circ}$, is also expected to provide
our results
closest to the original results of Mineshige et al.\ (1994a), since we use the
same outer radius, but a somewhat smaller inner radius and the difference between the
Newtonian and pseudo-Newtonian potentials are $< 10\%$ when $r_N = 20$.  
Indeed, for all these simulations $\beta = 1.65 \pm 0.04$,
 in rather good agreement with the earlier non-relativistic work.
The effect of varying the inclination angle is very small here,  since the GR
effects are weak.   This relative lack of GR effects is because the closest emission 
now comes from 
$20 r_g$; this implied  we
did not need to compute as many intermediate angle cases.

It is also instructive to compare these Cases 28--32 with the second set (10--18), in that
the ratios of $r_1/r_N$ are very similar (5 against 5.33).  Our earlier work (WX) argued that the
ratio of the outer to the inner edge of the unstable zone was a key determinant of the PSD slope,
and indeed the slopes for $i = 0$ are not too dissimilar ($1.55$ vs.\ $1.68$).
Furthermore, the gravitational red-shift acting more strongly  in Case 10 than in Case 28 could 
explain the bulk of that modest difference.  But once the viewing angle departs significantly
from face-on, the GR induced differences become quite substantial indeed,
and we conclude that the ratio of inner to outer zone radii is not a dominating variable.
Rather, what is most important is the breadth of the zone of instability, as long as the
inner boundary thereof is not far from the marginally stable orbit.

Cases 33--37 illustrate the effects of different prescriptions for propagating
the instability within the disk.  The standard models of Mineshige et al. (1994a, b) utilized
a rough presciption for differential rotation, whereby if the $(i,j)$ zone went
unstable, its dumped mass was equally transferred to the $(i+1,j+1-I),(i+1,j+2-I)$ 
and $(i+1,j+3-I)$ zones.  They typically considered $I = 1$, as have we throughout
the simulations so far; the value in the Rot.\ column of Table 1 is the value
of $I$ used in the simulations. Mineshige et al. (1994a) also considered
a case where $I=0$ and $I=2$ and found little difference in results.   No
differential rotation, other than that enforced by the quasi-Keplerian
azimuthal velocity (Eqn. 3), is modeled by $I=0$, and we have considered this in
Cases 33 and 36, which are the analogs to Cases 2 and 4 for $I=1$.  We find modest,
and consistent deviations, with $\Delta \beta = 0.05$ for Case 33 vs.\ Case 2 ($i = 30^{\circ}$)
and $\Delta \beta = 0.06$ for Case 36 vs.\ Case 4 ($i = 60^{\circ}$).  When we
consider larger amounts of ``differential rotation'', with $I = 3$ and $I =5$,
 the results
differ hardly at all from the ``rigid rotation'' case for lower inclination angle,
but the deviations increase with $I$, up to a total of $\Delta \beta = 0.15$
between $I = 1$ and $I=5$ for $i=60^{\circ}$.  While these changes are not
that dramatic,
we again find greater GR effects at higher viewing angles.
Note that in all these cases the actual velocity field is Keplerian and the 
differences as to which zones are triggered do not directly affect the Doppler
boosting calculations; however, different choices for $I$ do determine which
zones are included in the ``avalanche'' and therefore can modify the
total energy released.

We also briefly considered what would happen if an unstable zone
shared its dumped mass with zones not just inward of it, but also
to its sides and outward.  The rationale for doing so and the
way in which we implemented this modification
was discussed above, in \S 3.1.  Only two such simulations were
performed (Cases 38 and 39), and the column headed ``Diff.''
distinguishes them, in that the trigger is spread in all directions
as opposed to just inward, as is the case for every other
simulation.  As was the situation when GR effects
were ignored, we found a substantial
steepening of the PSD, with $\Delta \beta = 0.28$ for
$i = 30^{\circ}$ and $\Delta \beta = 0.43$ for
$i = 60^{\circ}$.  We can understand this result because
such a spreading of the triggering would encourage more  
longer-lived flares concentrated at the outer portion of
the disk, putting relatively more power in the
lower frequency bins.    We note that we did not
extend the approach of Mineshige et al.\ (1994b; cf. \S 2.1)
wherein
slow accretion always occurred, independent of whether any
zone was unstable in a given timestep, to include GR effects.
Our non-relativistic simulations of this situation agreed
well with theirs, and so we expect the modifications produced
by GR as discussed above to act in concert with those produced by simulations including
weak infall at a constant rate.

Our final set of simulations, Cases 40--43, examines the influence
of the size of the trigger.  Our earlier non-relativistic
simulations had considered a substantial number of variations
in the relative size of the trigger (the dumped mass, $m_d$)
and the initial deviation from the critical mass in each zone.
Those simulations (WX) found only very small changes in the PSDs
produced after the simulations were run for long times, although 
early-time light curves could be very different.  So we expected
little dependence on $m_d$ even when GR effects were included, and
this turns out to be so.  While we chose $m_d = 2$ for all other
cases, here we considered an order of magnitude range, $m_d = [0.5,
5.0]$.  The total range in the PSD slope was only 0.07 for 
$i = 30^{\circ}$ and 0.09 for $i=60^{\circ}$; in both cases the larger
dumped masses ($m_d = 5.0$) produce steeper power spectra, while the
smaller triggers ($m_d = 0.5$) yield shallower slopes.  

As is clear from Figs.\ 2--4, in many cases the PSD is not perfectly described
as a single power-law, even over the limited region in which
we calculate our $\beta$ values.  The lowest frequencies, where the
PSD rises,  must
clearly be excluded in a computation of an approximate power-law.
 The highest frequencies plotted exclude a
unphysical turnaround and thereby correspond to a cut-off below the Nyquist
limit.  The slight hump in the PSD (around $f = 100$) is more
pronounced at higher inclination angles (cf.\ Fig.\ 4), and is clearly induced
by GR effects.  Oscillations in the PSD at high frequencies are
also sometimes observed (e.g.\ Figs. 2 and 4), perhaps 
indicating a sort of QPO behavior at the
highest frequencies, though the overall slope is only negligibly
affected by these deviations from a simple power-law.

\section{Conclusions}

Our extension of attempts to produce SOC situations on accretion
disks around BHs by including GR effects have shown substantial
variations in detailed light curves and in the power spectrum
of the fluctuations.  Strong dependences on viewing angle
are noted and the details thereof differ from the preliminary
estimates by Abramowicz and Bao (1994), particularly at lines-of-sight
very close to the disk (though we note that in reality,
such extreme viewing angles are unlikely, as they could be 
obscured by the wider
outer regions of the disks, or, in the case of AGN, surrounding tori).  
Dramatic differences in the slope of
the PSD are found when different inner and outer radii for the
unstable disk region are considered.  This critical result is independent
of viewing angle and propagation effects, as can be seen from Table 1
for Cases 1, 10, 19 and 28. 
The PSD slope actually becomes flatter than is ever observed if the outer radius of
the unstable zone is too large ($\sim 100 r_g$ for the inner edge fixed
at the marginally stable orbit).  Unsurprisingly, GR effects
become unimportant if the inner boundary of the unstable region
ends at radii substantially greater than $6GM/c^2$.  Very large 
changes in the PSDs are also engendered by different assumptions
about how the triggering spreads through the disk.  Nonetheless,
several other parameters induce only minor changes in the long-term
light curves and power spectra:  modifications in the nominal rotation
prescription; changes to the mass dumped from an unstable zone;
and the size of the initial deviations from the critical state.

Given these many variations, it is only fair to conclude that self-organized
criticality, in its {\it strictest sense}, is {\it not} produced, even in these 
highly idealized accretion disk models.
The fundamental reason for this is that the basic
 units of energy release are no longer the same; rather, the amount of energy 
released deeper
in the gravitational well is more substantial for a given mass traversing a 
given distance.  (This effect is independent of whether uniform or logarithmically
spaced radial zones are used in the computations.)  When GR effects are included, the scale-free
nature required of an SOC process is further disrupted; the existence
of a physical scale ($r_g$) in the potential breaks a fundamental symmetry
really necessary for a complete SOC situation to be realized. Variations in the slope
of the PSD can be understood by considering which portions of the disk
release relatively more energy and by how the observations are affected
by Doppler shifts, light bending and focusing.  In addition, we saw that
non-power law features can be imposed on the observed PSDs with the proper inclusion
of GR effects.

 Although a more realistic physical model might
modify the statement concerning concentrated energy release made in 
the previous paragraph, it probably would not do so.
For example, consider magnetic flares based on reconnection events,
the basic physical background, explicit in the models of Lu \& Hamilton (1991)
and indirectly invoked by Mineshige et al. (1994a):  if the energy
density in the magnetic fields is
close to equipartition with the pressure in the matter, then higher energy releases 
would still be expected closer to the inner edge of the disk.  (In the case
of solar flares explicitly considered by Lu \& Hamilton, the assumption that
the reconnection process is nearly scale-free is more nearly viable
than in the case of accretion disks.)
 Therefore, even when the disk is operating 
right at the ``hairy edge'' of stability, so that flares of many different sizes
 can indeed be randomly 
triggered by a single infalling mass unit (or reconnection event) at the outer edge of
the unstable portion of the disk, thus implying that  an {\it approximate} SOC 
state exists, we do
not see all the properties of a full self-organized criticality.  This is
true even when the variations induced by GR on the propagation of the
emitted photons are ignored.  We stress that the key
observables, i.e., the light curves and PSD, still depend on the size and
location of the region over which the instability
operates; therefore, they  are {\it not completely independent of boundary conditions}, or other
small details, as  would be 
the case for a system that is in a {\it full-fledged} SOC state.  

Finally, GR effects certainly
make significant differences in what is observed, although, since they
involve the propagation of photons after leaving the disk, those externally
produced changes should not additionally count against a claim that SOC has
been established.
Nonetheless, it is intriguing that these models typically reproduce variability
 power spectrum slopes
between $-1.0$ and $-2.0$, as this also seems to span the range of observed PSD slopes seen
in both galactic and extragalactic variable sources.  Therefore, the importance
of quasi-SOC situations remains
an intriguing possibility, and more careful physical models for the triggering
of flares of a variety of strengths within and above accretion disks are well worth studying.

\par
\vspace{1pc}\par
We thank the anonymous referee for suggesting several
important corrections and clarifications.  PJW is most grateful 
for hospitality at the Department of Astrophysical Sciences 
of Princeton University where the writing of this paper was completed.
This work was supported by NASA Grant NAG 5-3098 and by Research Program Enhancement
funds at Georgia State University.

\newpage

\section*{References}
\small

\re
Abramowicz, M.A., Bao, G.\ 1994, PASJ 46, 523 

\re
Abramowicz, M.A., Bao, G., Lanza, A., Zhang, X.-H.\ 1989,
in Proc.\ 23rd ESLAB Symp. on Two Topics in X-Ray Astronomy
ed J.\ Hunt, B.\ Battrick. ESA Sp-296, (European Space Agency), p871

\re
Abramowicz, M.A., Bao, G., Lanza, A., Zhang, X.-H.\ 1991, A\&A 245, 454

\re
Abramowicz, M.A., Bao, G., Larsson, S., Wiita, P.J.\  1997,
ApJ 489, 819

\re
Abramowicz, M.A., Lanza, A., Spiegel, E.A., 
Szuszkiewicz, E. 1992, Nature 356, 41

\re
Artimova, I.V., Bj\"ornsson, G.,  Novikov, I.D.\ 1996, ApJ 461, 565

\re
Bak, P., Tang, C., Wiesenfeld, K.\ 1987, 
Phys.\ Rev.\ Lett.\  59, 381

\re
Bak, P., Tang, C., Wiesenfeld, K.\ 1988,
Phys.\ Rev. A.\  38, 364

\re
Bao, G., Abramowicz, M.A.\ 1996, ApJ 465, 646

\re
Bao, G., Hadrava, P., Ostgaard, E.\ 1994, ApJ 435, 55

\re
Bao, G., Hadrava, P., Wiita, P.J.,  Xiong, Y.\ 1997, 
ApJ 487, 142

\re
Bao, G., Wiita, P.J.\  1997, ApJ 485, 136
\re
Bao, G.,  Wiita, P.J.\  1999, ApJ 519, 80
\re
Bao, G., Wiita, P.J., Hadrava, P.\ 1996,
Phys.\ Rev.\ Lett.\  77, 12

\re
Bao, G., Wiita, P.J.,  Hadrava, P.\ 1998,
ApJ, 504, 58

\re
Bracco, A., Provenzale, A., Spiegel, E.A., Yecko, P.\ 1999, 
in Theory of Black Hole Accretion Disks, ed M.A.\ Abramowicz, 
G.\ Bj\"ornsson, J.E.\ Pringle, 
(Cambridge University Press, Cambridge) p274
\re
Chakrabarti, S.K.\ 1996, Phys.\ Reports  266, 229

\re
Chakrabarti, S.K., Wiita, P.J.\ 1993,  ApJ, 411, 602

\re
Chen, X., Abramowicz, M.A., Lasota, J.-P., Narayan, R.,  Yi, I.\
1995, ApJ 443, L61

\re
Dennis, B.R.\ 1985, Solar Physics 100, 465

\re
Ichimaru, S. 1977, ApJ, 214, 840

\re
Lawrence, A., Papadakis, I.\  1993, ApJ 414, L85

\re
Lightman, A.P., Eardley, D.M. 1974, ApJ 187, L1
\re
Lu, E.T.\ 1995, ApJ 446, L109

\re
Lu, E.T., Hamilton, R.J.\ 1991, ApJ 380, L89

\re
Mangalam, A.V.,  Wiita, P.J.\ 1993, ApJ 406, 420

\re
Manmoto, T., Takeuchi, M., Mineshige, S, Matsumoto, R., Negoro, H.\ 1996, ApJ 464, L135

\re
Mineshige, S., Negoro, H.\ 1999, in High Energy Processes in Accreting Black
Holes, ASP Conf. Ser. 161, ed. J. Poutonen, R. Svensson (Astr. Soc. Pacific, San
Francisco) p113

\re
Mineshige, S., Ouchi, N.B., Nishimori, H.\ 1994a, PASJ 46, 97

\re
Mineshige, S., Takeuchi, M., Nishimori, H.\ 1994b, ApJ 435, L125

\re
Miyamoto, S., Iga, S., Kitamoto, S.,  Kamado, Y.\ 1993, ApJ 403, L39

\re
O'Brien, K., Wu, L.,  Nagel, S.R.\ 1991, Phys.\ Rev.\ A\
43, 2052

\re
Paczy\'nski, B., Wiita, P.J.\ 1980, A\&A, 88, 23

\re
Parker, E.N. 1988, ApJ,  330, 474

\re
Press, W.H. 1978, Comments Ap.\ 7, 103

\re
Sawada, K., Matsuda, T., Hachisu, I.\ 1986, MNRAS 221, 679

\re
Shakura, N.I., Sunyaev, R.A.\ 1973, A\&A 24, 337

\re
Stern, B.E.,  Svensson, R.\ 1996,  ApJ 469, L109

\re
Takeuchi, M., Mineshige, S.\ 1997 ApJ, 486, 160

\re
Takeuchi, M., Mineshige, S., Negoro, H.\ 1995, PASJ 47, 617

\re
Vaughn, B.A., Nowak, M.A.\ 1997, ApJ 474, L43

\re
Wiita, P.J., Miller, H.R., Carini, M.T., Rosen, A. 1991,
in Structure
and Emission Properties of Accretion Disks, ed C.\ Bertout, 
S.\ Collin-Souffrin, J.P.\ Lasota, J.\ Tran Thanh Van, 
6th I.A.P.\ Astrophysics Meeting / 
I.A.U.\ Colloq.\  129,  (Editions Fronti\'eres, Gif-sur-Yvette) p557

\re
Wiita, P.J., Xiong, Y.\ 1999, in  Theory of Black
Hole Accretion Disks, ed M.A.\ Abramowicz, G.\ Bj\"ornsson,  J.E.\ Pringle,
(Cambridge University Press, Cambridge) p274 (WX).

\re
Yonehara, A., Mineshige, S., Welsh, W.F.\ 1997, ApJ 486, 388

\re
Zhang, X.-H., Bao, G.\ 1991, A\&A 246, 21

\label{last}

\newpage

\begin{table}[t]
\small
\begin{center}
Table~1.\hspace{4pt}Parameters and Results of Selected Simulations.\\
\end{center}
\vspace{6pt}
\begin{tabular*}{\columnwidth}{@{\hspace{\tabcolsep}
\extracolsep{\fill}}p{1pc}cccccccc} 
\hline\hline\\[-6pt]
Case &$r_{in}$ &$r_{out}$ & $i$ &  Rot. & Diff.&	m$_d$& $\beta$ \\
[4pt]\hline \\[-6pt]
1  &  6 & 19 & 0 &  1 &  in  &  2 & 1.69  \\
2  &  6 & 19 & 30 &  1 &  in  &  2 & 1.55  \\
3  &  6 & 19 & 45 &  1 &  in  &  2 & 1.43  \\
4  &  6 & 19 & 60 &  1 &  in  &  2 & 1.35  \\
5  &  6 & 19 & 65 &  1 &  in  &  2 & 1.35  \\
6  &  6 & 19 & 70 &  1 &  in  &  2 & 1.34  \\
7  &  6 & 19 & 74 &  1 &  in  &  2 & 1.36  \\
8  &  6 & 19 & 80 &  1 &  in  &  2 & 1.43  \\
9  &  6 & 19 & 85 &  1 &  in  &  2 & 1.56  \\
&&&&&&& \\
10 &  6 & 32 &  0 &  1 &  in  &  2 & 1.55  \\
11 &  6 & 32 & 30 &  1 &  in  &  2 & 1.38  \\
12 &  6 & 32 & 45 &  1 &  in  &  2 & 1.33  \\
13 &  6 & 32 & 60 &  1 &  in  &  2 & 1.33  \\
14 &  6 & 32 & 65 &  1 &  in  &  2 & 1.23  \\
15 &  6 & 32 & 70 &  1 &  in  &  2 & 1.20  \\
16 &  6 & 32 & 74 &  1 &  in  &  2 & 1.21  \\
17 &  6 & 32 & 80 &  1 &  in  &  2 & 1.26  \\
18 &  6 & 32 & 85 &  1 &  in  &  2 & 1.30  \\
&&&&&&& \\
19 &  6 & 100&  0 &  1 &  in  &  2 & 1.02  \\
20 &  6 & 100&  30 &  1 &  in  &  2 & 0.93  \\
21 &  6 & 100&  45 &  1 &  in  &  2 & 0.86  \\  
22 &  6 & 100&  60 &  1 &  in  &  2 & 0.83  \\
23 &  6 & 100&  65 &  1 &  in  &  2 & 0.79  \\

\end{tabular*}
\end{table}
\newpage
\begin{table}[t]
\small
\begin{center}
Table~1, continued.\hspace{4pt}Parameters and Results of Selected Simulations.\\
\end{center}
\vspace{6pt}
\begin{tabular*}{\columnwidth}{@{\hspace{\tabcolsep}
\extracolsep{\fill}}p{1pc}cccccccc} 
\hline\hline\\[-6pt]

24 &  6 & 100&  70 &  1 &  in  &  2 & 0.79  \\
25 &  6 & 100&  74 &  1 &  in  &  2 & 0.76  \\
26 &  6 & 100&  80 &  1 &  in  &  2 & 0.75  \\
27 &  6 & 100&  85 &  1 &  in  &  2 & 0.77  \\
&&&&&&&\\
28 & 20 &100 &  0 & 1 & in & 2 & 1.68 \\
29 & 20 &100 &  30 & 1 & in & 2 & 1.69 \\
30 & 20 &100 &  60 & 1 & in & 2 & 1.69 \\ 
31 & 20 &100 &  70 & 1 & in & 2 & 1.64 \\
32 & 20 &100 &  85 & 1 & in & 2 & 1.62 \\ 
&&&&&&&\\
33 & 6 & 19 &  30 & 0 & in  & 2 & 1.60 \\
34 & 6 & 19 &  30 & 3 & in  & 2 & 1.59 \\
35 & 6 & 19 &  30 & 5 & in  & 2 & 1.59 \\
36 & 6 & 19 &  60 & 0 & in  & 2 & 1.41 \\
36 & 6 & 19 &  60 & 3 & in  & 2 & 1.46 \\
37 & 6 & 19 &  60 & 5 & in  & 2 & 1.51 \\
&&&&&&&\\
38 & 6 & 19 &  30 & 1 & all  & 2 & 1.83 \\ 
39 & 6 & 19 &  60 & 1 & all  & 2 & 1.78 \\
&&&&&&&\\
40 & 6 & 19 &  30 & 1 & in  & 0.5 & 1.58 \\
41 & 6 & 19 &  30 & 1 & in  & 5 & 1.51 \\
42 & 6 & 19 &  60 & 1 & in  & 0.5 & 1.43 \\
43 & 6 & 19 &  60 & 1 & in  & 5 & 1.34 \\ 

\hline
\end{tabular*}
\end{table}


\newpage

\begin{figure}
\epsffile{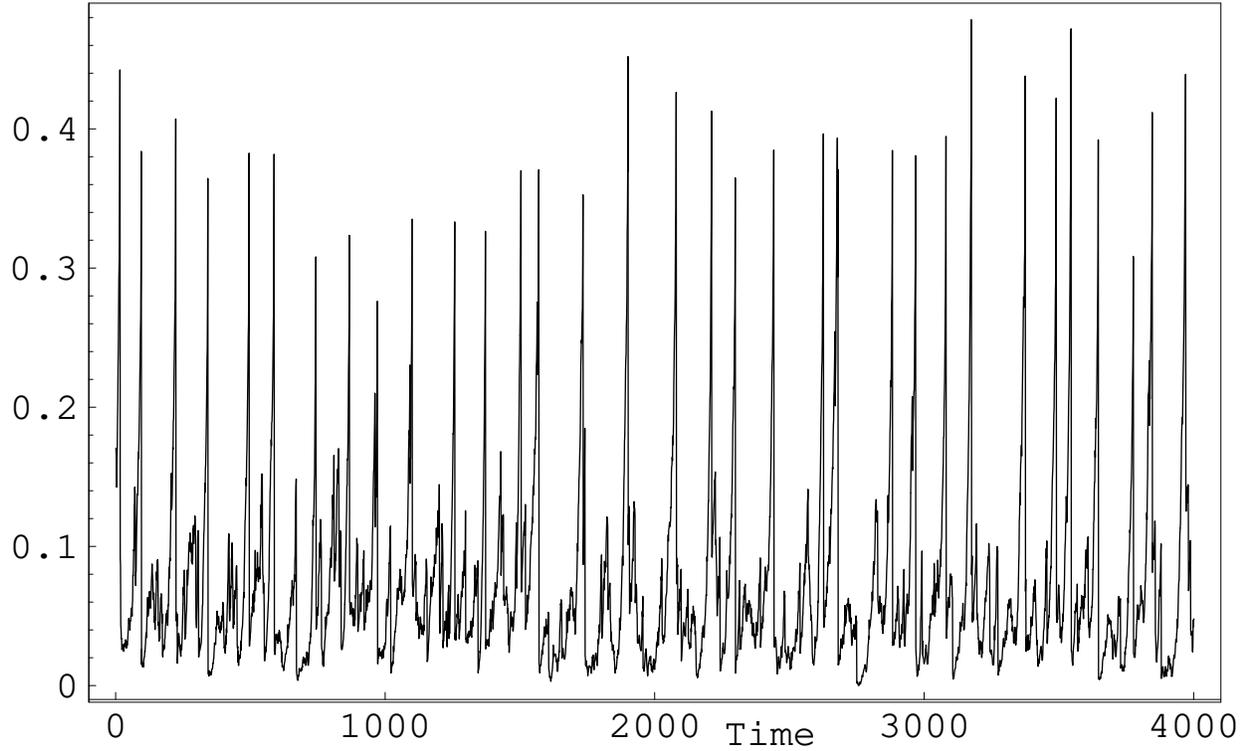}
\epsffile{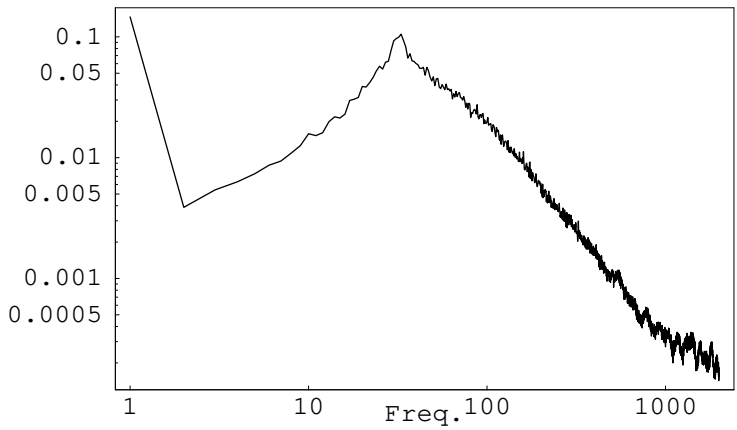}
\caption{(a) A portion of the light curve for Case 2, with $r_{in} =6, r_{out} =19,
i = 30^{\circ}$.  This segment is chosen to be short enough
to allow fine structure to be seen and is also selected from a piece of the
entire simulation far enough developed so that any initial condition effects
have dissipated. (b) The power spectrum density for Case 2; this curve is
a mean of PSDs computed from eight equal segments of the light-curve, and
the value of its slope, $\beta$, quoted in Table 1, is fit to this curve to 
frequencies
between 50 and 2048.}
\end{figure}

\newpage
\begin{figure}
\epsffile{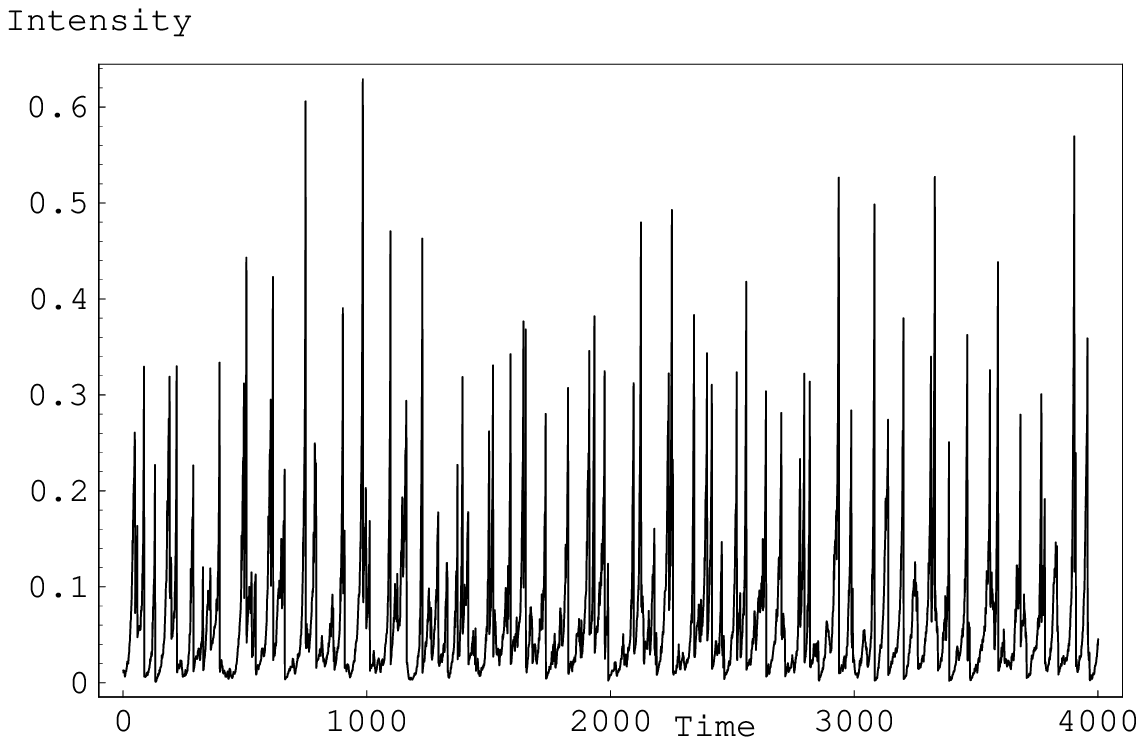}
\epsffile{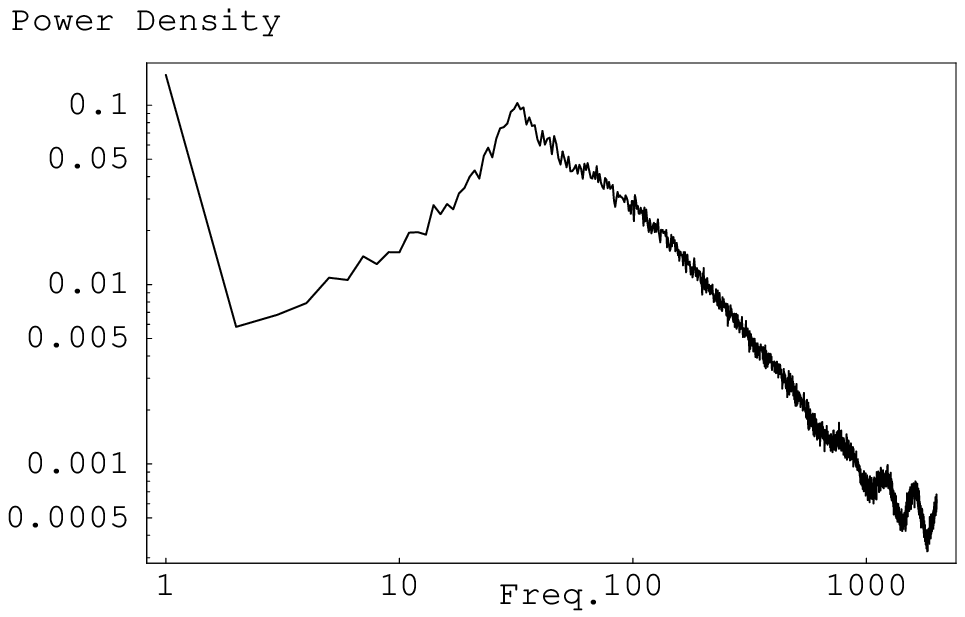}
\caption{As in Fig.\ 1 for Case 11, with $r_{in} = 6, r_{out} = 32, 
i = 30^{\circ}$. Note that these figures
all have different y-axis scales.}
\end{figure}

\newpage

\begin{figure}
\epsffile{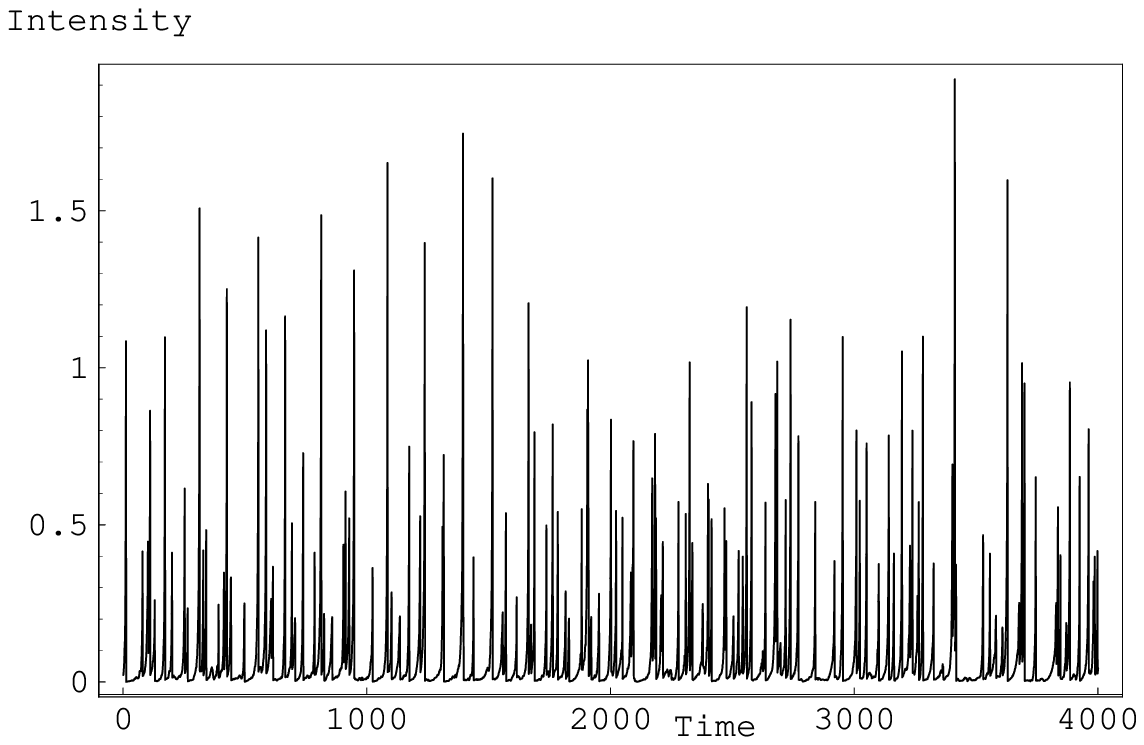}
\epsffile{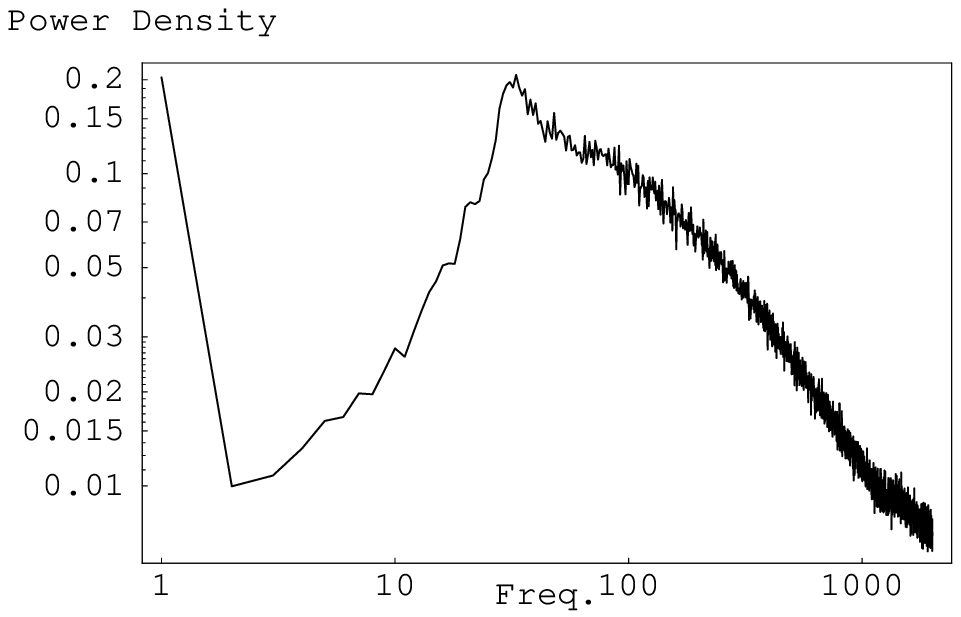}
\caption{As in Fig.\ 1 for Case 20, with $r_{in} =6, r_{out} =100; 
i=30^{\circ}$.}
\end{figure}

\newpage
\begin{figure}
\epsffile{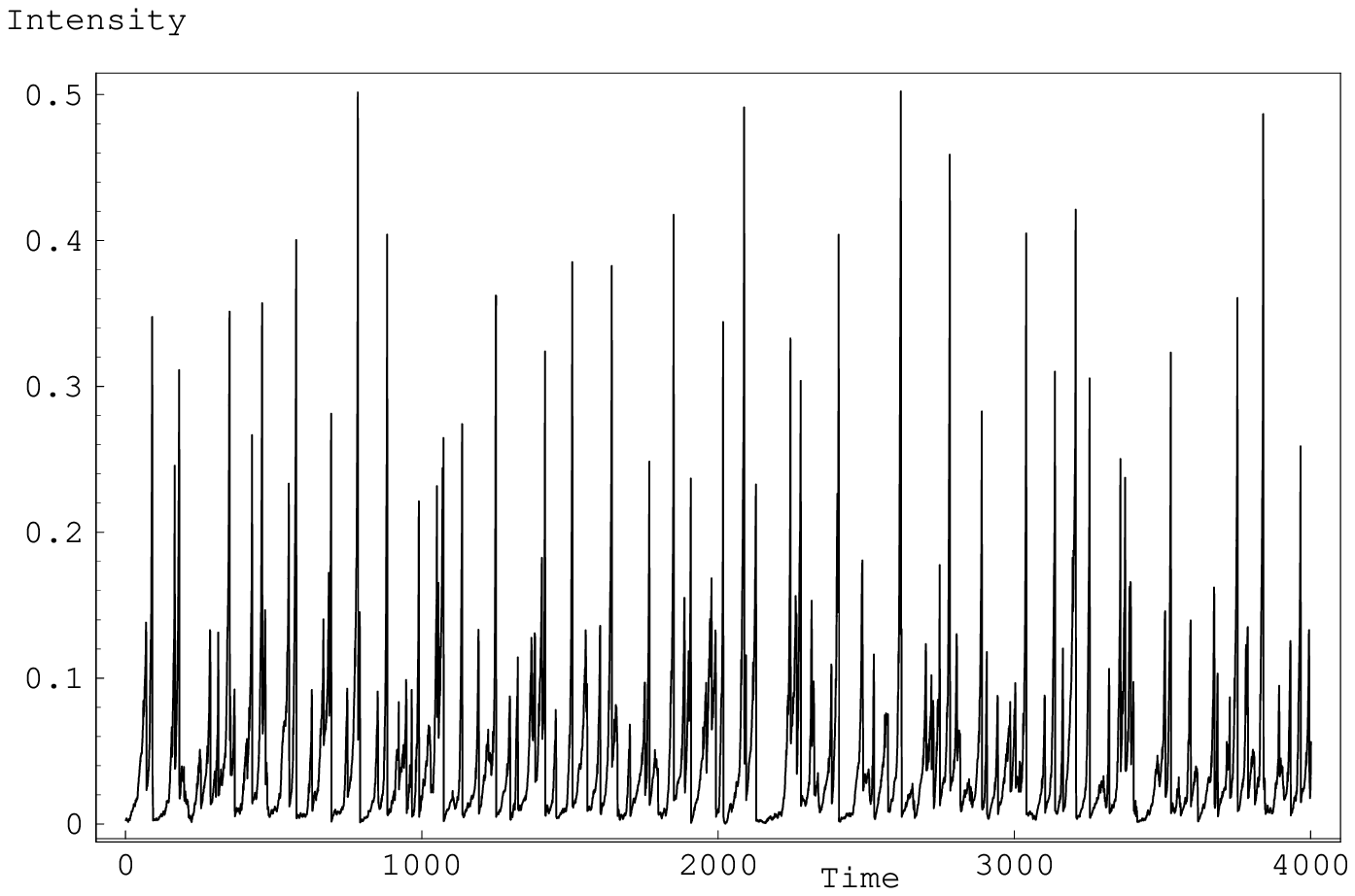}
\epsffile{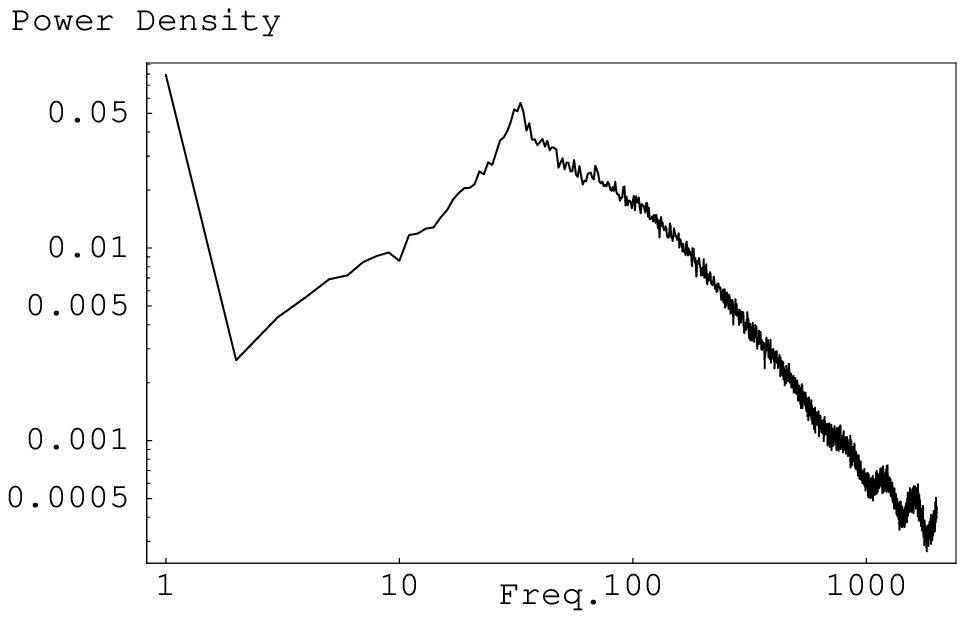}
\caption{As in Fig.\ 1 for Case 13, with $r_{in}=6, r_{out}=32, 
i=60^{\circ}$.}
\end{figure}

\end{document}